\documentclass[twocolumn,epsfig,pre]{revtex4}

\usepackage{graphics}
\usepackage{graphicx}
\usepackage{epstopdf}
\usepackage{amssymb}
\usepackage{amsmath}
\usepackage{hyperref}
\usepackage{latexsym}
\usepackage{color}

\begin{document}
	
	\renewcommand{\thefootnote}{\fnsymbol{footnote}}
	\renewcommand{\theequation}{\arabic{section}.\arabic{equation}}
	
	\title{Do chain topology and polydispersity affect the two-stage heteropolymer coil-globule transition?}

	\author{Thoudam Vilip Singh}
	\email{thoudamsinghsingh@gmail.com}	
	\author{Lenin S.~Shagolsem}
	\email{leninshagolsem@nitmanipur.ac.in}
	\affiliation{Department of Physics, National Institute of Technology Manipur, Imphal, India} 
	
	\date{\today}

\begin{abstract}

\noindent The thermodynamic behavior of collapse transition in a fully flexible coarse-grained model of energy polydisperse polymer (EPP), a statistical model of random heteropolymer, is investigated in an implicit solvent by means of molecular dynamics (MD) simulations. Each monomer has interaction energy, $\varepsilon_i$, randomly drawn from a Gaussian distribution, and is characterised by polydispersity index, $\delta$ = standard deviation/mean, where the mean is fixed at $\langle \varepsilon \rangle$ = 2.5. Polymers of different chain topologies assume an expanded coil conformation at high temperature, and undergo a melting transition called a coil-globule transition when temperature is lowered. They collapse into molten globules. Further decrease in temperature results in a liquid-solid transition called a freezing transition, thus, creating crystallite structures at very low temperatures. The current study investigates the effect of chain topology and energy polydispersity in this regard from thermodynamic point of view. 

\end{abstract}

\maketitle



Investigating the coil-globule transition in coarse-grained heteropolymer models is crucial for understanding fundamental physical phenomena such as protein folding. Consequently, extensive theoretical and experimental research has been carried out in this field.\cite{Sherman,Tanaka0,Kita,Frerix,Xuuu,Nakata} The $\theta$ or transition point of a linear chain as it undergoes a collpase transition can be determined using swelling coefficient, Steinhauser's approach, etc.\cite{Vilip2021,Vilip2022} For alternative topologies like unknotted rings and trefoil knots, $g$-factor serves as a useful metric for locating the relative transition points.\cite{Vilip2023} This Letter focuses on the thermodynamic aspects of the coil-globule transition. In linear polymer chains, two distinct phase transitions have been reported: (a) athe coil-globule transition, a melting-like process that occurs as temperature decreases, and (b) the subsequent freezing (liquid-solid) transition at even lower temperatures, marked by an increase in polymer density. These states are often referred to as the crumpled globule and compact globule, respectively.\cite{Wu1995,Yu1992} The heat capacity at constant volume is computed as $C_V = \text{d}U/\text{d}T$ where $U(T)$ represents the internal energy of the polymer system. Peaks in the $C_V(T)$ curves correspond to phase transitions like the coil-globule and the liquid-solid transitions. Previous studies have employed techniques like Wang-Landau sampling and various polymer models, including bond-fluctuation and off-lattice models, to investigate these transitions.\cite{Seaton2010,Paul2007,Rampf2005,Parsons2006-1,Parsons2006-2,Seaton2009} However, most of these studies have primarily focused on single, flexible homopolymers with linear topology but varying chain lengths. Wust and co-workers examined a simple and linear heteropolymer lattice model of protein consisting of two monomer species (hydrophobic and polar) using Wang-Landau Monte Carlo simulations.\cite{Wust2011}

Here, we present results of MD simulations for a fully flexible heteropolymer model with varying topologies and chain lengths, comparing them to their homopolymer counterparts. Our generic heteropolymer model is based on an independent interaction framework, where the number of monomer species equals the total number of monomers. Each monomer possesses a randomly assigned interaction energy,  $\varepsilon_i$, drawn randomly from a Gaussian distribution. These heteropolymers, referred to as EPP chains (or simply heteropolymer here), serve as statistical representation of biopolymers such as intrisically disordered proteins (IDP).\cite{Pande-Ch6, Shakhnovich-Ch6, Vilip2022, Vilip2023} We have examined the impact of different functional forms and variance in energy distribution on the coil-globule transition using key physical parameters related to radius of gyration, $\langle R_g^2 \rangle = \langle \frac{1}{N} \sum_{i=1}^{N}\left( \textbf{r}_i - \textbf{r}_{cm} \right)^2 \rangle$, like expansion factor, and $g$-factor.\cite{Vilip2022,Vilip2023} Here, $\textbf{r}_i$ and $\textbf{r}_{cm}$ denote the monomer position and that of centre of mass of the chain respectively. In this study, we investigate whether a two-stage phase transition occurs in our heteropolymer model and assess the influence of chain topology from a thermodynamic perspective. Additionally, we analyze and report the effects of interaction energy polydispersity.





\begin{figure*}[ht]
	\begin{center}
		\includegraphics*[width=1\textwidth]{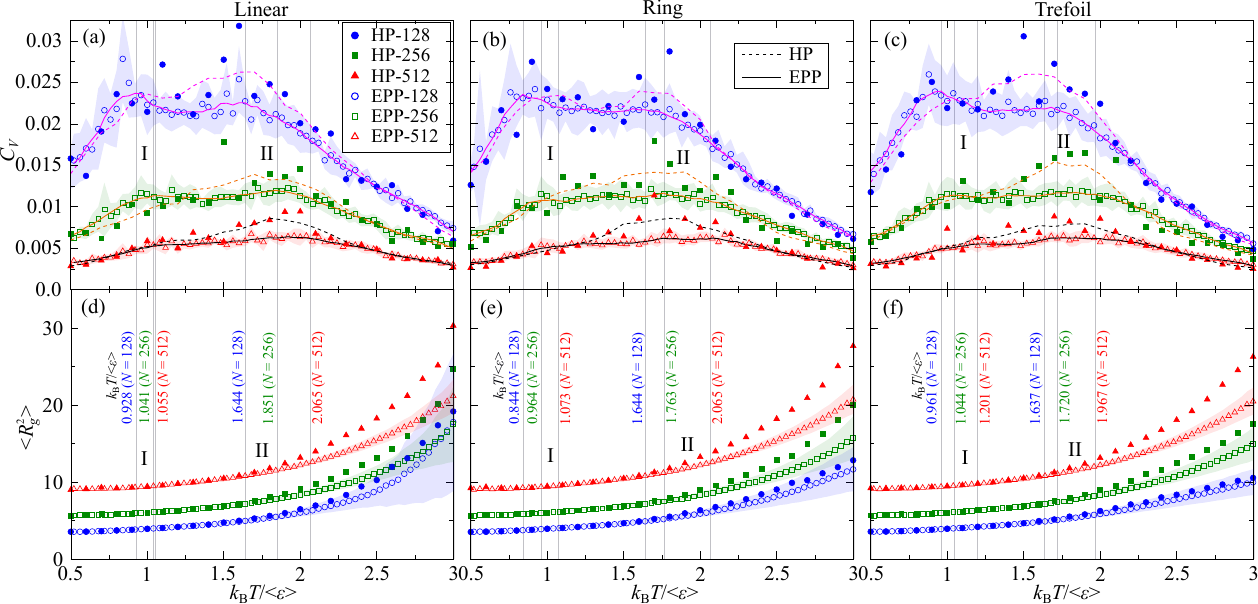}
		\caption{Plots of heat capacity at constant volume, $C_V$, as a function of temperature (rescaled by mean energy, $\langle \varepsilon \rangle$) for energy polydisperse polymers and homopolymers are shown in (a)-(c) for linear, ring and trefoil respectively at different chain lengths, $128 \le N \le 512$. The corresponding temperature dependence of $\langle R_g^2 \rangle$ for different chain topologies are displayed in (d)-(f). Filled symbols indicate homopolymers, and open symbols that of the heteropolymers. The vertical lines denote the freezing transition temperatures (near peak I), and melting transition or coil-globule transition temperatures (near peak II). All systems displayed here have a polydispersity index of $\delta$ = 9\%.}
		\label{fig: Cv-Gd9-all-topo}	
	\end{center}
\end{figure*}

\begin{figure}[]
	\begin{center}
		\includegraphics*[width=0.47\textwidth]{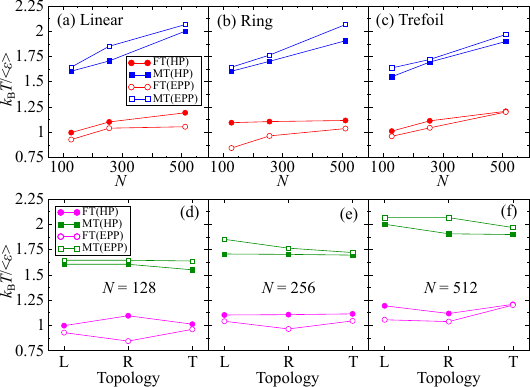}
		\caption{Plots highlighting the effect of chain length, $N$, on the melting (MT) and freezing transition (FT) temperatures in figure (a)-(c) for linear, ring and trefoil respectively. In figure (d)-(f), the transition temperature dependence on chain topology is displayed for different chain lengths, $N$ = 128$-$512. L, R and T implies linear chain, simple ring and knotted trefoil respectively. Filled symbols indicate homopolymers (HP), and open symbols that of the heteropolymers (EPP). All systems displayed here have polydispersity index of $\delta$ = 9\%.}
		\label{fig: N-vs-transition}	
	\end{center}
\end{figure}

\begin{figure*}[ht]
	\begin{center}
		\includegraphics*[width=1\textwidth]{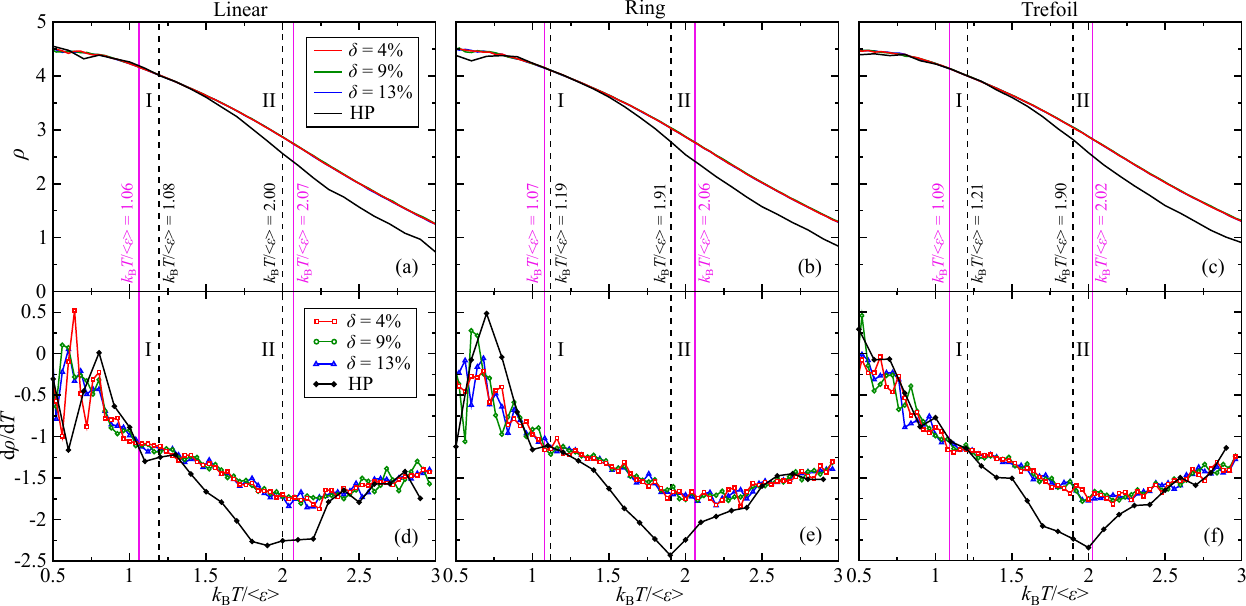}
		\caption{Plots of number density, $\rho$, as a function of rescaled temperature, $k_{\rm B}T/\langle \varepsilon \rangle$, for (a) linear, (b) ring and (c) trefoil polymers at different values of $\delta$. Their respective temperature derivaties are shown in (d)-(f) respectively. The chain length of the systems are fixed at $N$ = 512.}
		\label{fig: rho-delta}	
	\end{center}
\end{figure*}

\begin{figure}[ht]
	\begin{center}
		\includegraphics*[width=0.48\textwidth]{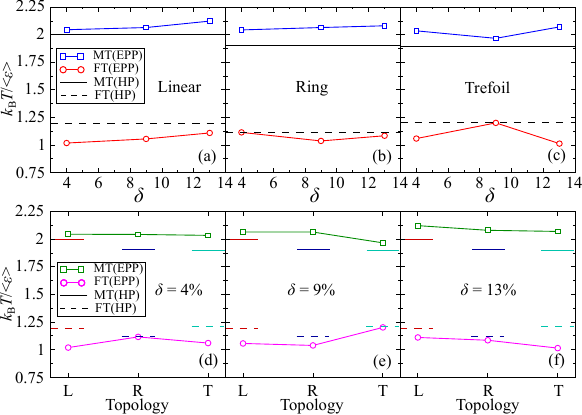}
		\caption{Plots highlighting the effect of polydispersity, $\delta$, on the melting (MT) and freezing transition (FT) temperatures in figure (a)-(c) for linear, ring and trefoil respectively. L, R and T implies linear chain, simple ring and knotted trefoil respectively. In figure (d)-(f), the transition temperature dependence on different chain topology is displayed for differen polydispersity index, $\delta$ = 4$-$13\%. The solid lines indicate melting transition for homopolymers, and the dashed lines indicate freezing transition for the heteropolymers. All systems displayed here have the same chain length, $N$ = 512.}
		\label{fig: delta-vs-transition}	
	\end{center}
\end{figure}


The heteropolymer chain is a coarse-grain bead-spring model of Kremer and Grest,\cite{kremer_JCP_92} where every monomer interacts via Lennard-Jones (LJ) potential. Throughout this Letter, every physical quantity is expressed in LJ reduced units,\cite{Frenkel,Allen} where $\sigma$ and $\varepsilon$ are the basic length and energy scales respectively. The LJ potential is cut-off and shifted to zero at $r=2.5\sigma$ ($\sigma$ being the monomer diameter). The monomer connectivity is assured via ﬁnitely extensible, nonlinear, elastic
(FENE) springs.\cite{grest_kramer_PRA_33} The spring constant is $k=30\varepsilon/\sigma^2$ and the maximum extension between two consecutive monomers along the chain is $r_0=1.5\sigma$.\cite{kremer_JCP_92, Kremer} The interaction energies of monomers, $\varepsilon_i$, are chosen randomly from a Gaussian distribution, 
\begin{equation}
P(\varepsilon_i)= 
\frac{1}{\sqrt{2 \pi \left( \text{SD} \right)^2}} \text{exp} \left[-\frac{\left(\varepsilon_i - \langle\varepsilon \rangle \right)^2}{4 \left( \text{SD} \right)^2} \right]~, 
\label{eqn: gaussian-dist}
\end{equation}
with SD as the standard deviation and $\langle \varepsilon \rangle = \left(\sum_{i=1}^{N} \varepsilon_i\right)/N$ as the mean. The mean is fixed at $\langle\varepsilon\rangle = 2.5$. By varying the SD, the energy polydispersity is introduced and its degree of polydispersity is quantified through polydispersity index $\delta~(={\rm SD}/{\rm mean})$.

Here, molecular dynamics (MD) simulations are employed using Langevin dynamics,\cite{Allen,Frenkel} and the equations of motion are integrated using velocity-Verlet scheme. A single polymer of a given topology is simulated in a cubic box ($L_x = L_y = L_z =$ 130$\sigma$) that is periodic in every direction. Three chain lengths, $N$ =  128, 256 and 512 are considered. The initial configuration is taken from an equilibrated polymer at an athermal condition which has been relaxed for enough time. These production runs last for $5 \times 10^6$ MD steps at the temperature range of $T$ = 1.1 to 7.5. The results reported here are obtained by averaging a maximum of 10 replicas, with each replica possessing a different set of interaction parameter, $\varepsilon_i$, from a Gaussian distribution. All the simulations are carried out using LAMMPS code.\cite{Plimpton}



To study an aspect of coil-globule transition in heteropolymers (EPP chains), the dependence of heat capacity at constant volume, $C_V(T)$, on rescaled temperature, $k_{\rm B}T/\langle \varepsilon \rangle$, for different topologies and chain lengths are studied at polydispersity ranging from $\delta$ = 4$-$13\%. In figure~\ref{fig: Cv-Gd9-all-topo}(a)-(c), we display the same at $\delta$ = 9\% for linear chain, simple ring, and trefoil knot respectively (See SI for $\delta$ = 4\% and 13\%). The corresponding variation of $\langle R_g^2 \rangle$ with $k_{\rm B}T/\langle \varepsilon \rangle$ are shown in figure~\ref{fig: Cv-Gd9-all-topo}(d)-(f). Here, $\langle R_g^2 \rangle = \langle \lambda_1^2 \rangle + \langle \lambda_2^2 \rangle + \langle \lambda_3^2 \rangle$ such that $\lambda_1^2 \le \lambda_2^2 \le \lambda_3^2$ are the eigenvalues of radius of gyration tensor.\\

A two-stage transition is observed in both the heteropolymers and the reference homopolymers, regardless of topology, chain length, and polydispersity index. At high temperatures, polymers of any given topology adopt an expanded coil conformation. As the temperature decreases, a melting transition occurs, wherein the polymer coil rapidly collapses into a molten globule, as indicated by the second peak (approximately in the regions marked as II). This process is commonly referred to as the coil-globule transition. Upon further lowering the temperature, a freezing transition takes place (approximately in the regions marked as I), analogous to a liquid-solid transition. The molten structure undergoes a transformation into a frozen globule, which is relatively more compact as reflected in figure~\ref{fig: Cv-Gd9-all-topo}(d)-(f). The vertical lines denote the transition points for EPP chains, corresponding to the peak positions of the average heat capacity curves obtained from 10 different replicas. With reference to these vertical lines, the values of $\langle R_g^2 \rangle$, in figure~\ref{fig: Cv-Gd9-all-topo}(d)-(f), are slightly larger in the vicinity of peak II compared to peak I, where they become nearly constant. The reorganisation of monomers within the liquid globule occurs in the intermediate region between the two peaks. Once the system reachs peak I as temperature decreases, the monomer positions remain effectively fixed. 

The dependence of the melting (MT) and freezing transitions (FT) on ahcin length $N$ is presented in figure~\ref{fig: N-vs-transition}(a)-(c) for linear chains, simple rings, and trefoil knots at a fixed polydispersity of $\delta$ = 9\%. In general, the transition temperatures for both MT and FT increase with increasing chain length, aligning with the coil-globule transition temperatures previously derived from expansion factors or swelling coefficients.\cite{Vilip2022} 
Another notable observation is that it takes less temperature to transition from molten to frozen globules in homopolymers than in heteropolymers. A notable observation is that homopolymers require a lower temperature to transition from molten to frozen globules compared to heteropolymers. Across all topologies and chain lengths, heteropolymers exhibit a larger temperature gap for this transition due to identity ordering,\cite{Vilip2022} wherein monomers with higher interaction energies migrate toward the center, while those with lower interaction energies remain at the periphery. This ordering effect is particularly pronounced in energy-polydisperse polymers with a uniform distribution. Figure~\ref{fig: N-vs-transition}(d)-(f) further illustrates the impact of topology on melting and freezing transitions across various chain lengths. A key and consistent finding is that the coil-globule transition temperature (or the melting point or the $\theta$-temperature) is the highest for linear chains, followed by simple rings, and the lowest for trefoil knots, for both heteropolymers and homopolymers. However, this trend does not hold for the freezing transition.

The number density, $\rho = N/\left(4\pi \langle R_g \rangle^3/3 \right)$, is plotted as a function of rescaled temperature, $k_{\rm B}T/\langle \varepsilon \rangle$, for different topologies within the polydispersity range $\delta$ = 4$-$13\% as shown in figure~\ref{fig: rho-delta}(a)-(c) for a fixed chain length, $N$ = 512. The corresponding temperature derivaties are presented in figure~\ref{fig: rho-delta}(d)-(f). As expected, the density decreases with increasing temperature, and the difference in the behavior of homopolymers and heteropolymers becomes more pronounced at higher temperatures. The melting transition (approximately in the regoins marked as II) occurs at a higher temperature for EPP chains relative to the homopolymers, as indicated by the vertical lines. This contrasts with our previous findings based on the expansion factor and the $g$-factor.\cite{Vilip2022,Vilip2023} A possible explanation is that during chain collapse, monomers with varying interaction energies more easily establish favorable contacts. Consequently, a higher temperature is required for a molten heteropolymer globule to expand into a coil, as the molten structure stabilized through identity ordering is more thermodynamically favorable than that of a homopolymer. The vertical solid lines represent the transition points for heteropolymers (averaged for $\delta$ = 4, 9 and 13\%), while the vertical dashed lines denote those of homopolymers. On the other hand, the freezing transition (approximately in the regions marked as I) occurs at a lower temperature for the heteropolymers relative to the reference homopolymers, a trend consistent across all topologies examined. Furthermore, both melting and freezing transitions are reflected in the temperature derivative of $\rho$ in in figure~\ref{fig: rho-delta}(d)-(f).

Figure~\ref{fig: delta-vs-transition}(a)-(c) illustrates the impact of energy polydispersity on the melting and freezing transitions for linear, ring, and trefoil polymer chains, respectively. In our previous study on linear heteropolymers, we observed that the $\theta$-temperature increases with higher values of $\delta$. A similar trend appears to hold in this case; however, the increase is relatively small, and the trend does not extend to trefoil knots. Further investigation is necessary to fully understand the influence of energy polydispersity on chain topology. An interesting observation is that, akin to the effect of chain length on transition temperatures, the temperature difference between the melting and freezing points is smaller for homopolymers than for energy polydisperse polymers. Figure~\ref{fig: delta-vs-transition}(d)-(f) presents the effect of topology on melting and freezing transition temperatures for $\delta$ = 4, 9 and 13\% respectively. Across all values of $\delta$, the coil-globule transition temperature is lowest for trefoil knots and highest for linear chains. However, this trend is not consistently observed in the freezing transition.


In conclusion, we present a heteropolymer model incorporating variations in chain topology, chain length, and polydispersity within the range of $\delta$ = 4$-$13\%. The thermodynamic behavior of the two-stage transition in fully flexible polymers is found to be universal. Regardless of topology, chain length, or polydispersity, the polymers remain in an expanded coil state at high temperatures before undergoing a coil-globule (melting) transition at the $\theta$-temperature. Upon further lowering the temperature, a freezing transition—analogous to a liquid-solid transition—occurs, leading to the formation of a highly compact and stable structure. TThe effect of chain length is observed in the increasing transition temperatures with longer chains. Notably, heteropolymer chains undergo melting at a higher temperature and freezing at a lower temperature compared to homopolymers, suggesting that an intermediate identity ordering occurs based on the interaction energy of the monomers.\cite{Vilip2022} As a result, the melting transition temperatures of energy polydisperse polymer (EPP) chains, determined thermodynamically, differ from those obtained using physical parameters involving  $\langle R_g \rangle$. Although heteropolymers have higher melting transition temperatures relative to the reference homopolymers, other results remain consistent: (a) the coil-globule transition temperature is highest for linear chains and lowest for trefoil knots, regardless of chain length and polydispersity, and (b) this trend is also observed in homopolymer systems. Among heteropolymers, the effect of polydispersity on transition behavior appears to be minimal. Future studies could explore the impact of different polydispersity distributions, such as uniform and exponential distributions, to gain a deeper understanding of their influence.



\begin{center}
	\textbf{AUTHOR INFORMATION}
\end{center}
\textbf{Corresponding Author} \par 
Lenin S.~Shagolsem -- {\it Department of Physics, National Institute of Technology Manipur, Imphal - 795004, India}; orcid.org/0000-0003-0836-7491; Email: lenin.shagolsem@nitmanipur.ac.in \\
\textbf{Authors} \par 
Thoudam Vilip Singh -- {\it Department of Physics, National Institute of Technology Manipur, Imphal - 795004, India}; orcid.org/0000-0003-4375-0402; Email: thoudamsinghsingh@gmail.com

\medskip
\textbf{Notes} \par
The authors declare no competing financial interest.

\medskip
\textbf{Data Availability} \par
The data that support the findings of this study are available from the corresponding author upon reasonable request.

\end{document}